\documentclass[%
 aip,
 amsmath,amssymb,
preprint,%
]{revtex4-1}

\DeclareSymbolFont{matha}{OML}{txmi}{m}{it}
\DeclareMathSymbol{\varv}{\mathord}{matha}{118}

\usepackage{graphicx}
\usepackage{dcolumn}
\usepackage{bm}

\usepackage[utf8]{inputenc}
\usepackage[T1]{fontenc}
\usepackage{mathptmx}
\usepackage{etoolbox}

\usepackage[T1]{fontenc}
\usepackage[utf8]{inputenc}
\usepackage[absolute,overlay]{textpos}
\usepackage{hyperref}
\usepackage{graphicx}
\usepackage{dcolumn}
\usepackage{bm}
\usepackage{xcolor}
\usepackage{siunitx}
\usepackage[all]{nowidow}

\newcommand{\blu}[1]{{\color{black}{#1}}}
\newcommand{\org}[1]{{\color{black}{#1}}}

\newcommand*{\citen}{}
\DeclareRobustCommand*{\citen}[1]{%
  \begingroup
    \romannumeral-`\x 
    \setcitestyle{numbers}%
    \cite{#1}%
  \endgroup
}

\makeatletter
\def\@email#1#2{%
 \endgroup
 \patchcmd{\titleblock@produce}
  {\frontmatter@RRAPformat}
  {\frontmatter@RRAPformat{\produce@RRAP{*#1\href{mailto:#2}{#2}}}\frontmatter@RRAPformat}
  {}{}
}%
\makeatother

\begin{document}

\preprint{AIP/123-QED}

\title[]{Coalescence of Surfactant-Laden Droplets}
\author{Soheil Arbabi}
\affiliation{Institute of Physics, Polish Academy of Sciences, Al. Lotnik\'ow 32/46, 02-668 Warsaw, Poland}
\author{Piotr Deuar}%
\affiliation{Institute of Physics, Polish Academy of Sciences, Al. Lotnik\'ow 32/46, 02-668 Warsaw, Poland}
\author{Mateusz Denys}
\affiliation{Institute of Physics, Polish Academy of Sciences, Al. Lotnik\'ow 32/46, 02-668 Warsaw, Poland}
\author{Rachid Bennacer}
\affiliation{Universit\'e Paris-Saclay, ENS Paris-Saclay, CNRS, LMPS, 4 Av. des Sciences, 91190 Gif-sur-Yvette, France}
\author{Zhizhao Che}
\affiliation{%
State Key Laboratory of Engines, Tianjin University, 300350 Tianjin, China
}%
\author{Panagiotis E. Theodorakis*}%
 \email{panos@ifpan.edu.pl}
\affiliation{Institute of Physics, Polish Academy of Sciences, Al. Lotnik\'ow 32/46, 02-668 Warsaw, Poland}

\date{\today}

\begin{abstract}
Droplet coalescence is an important process in nature and 
various technologies (\textit{e.g.} inkjet printing). 
Here, we unveil the surfactant mass-transport
mechanism and report on several major differences in
the coalescence of surfactant-laden droplets
as compared to pure water droplets
by means of molecular dynamics 
simulation of a coarse-grained model. 
Large scale changes to bridge growth dynamics
are identified, such as the lack of multiple thermally
excited precursors, attenuated collective excitations
after contact, slowing down in the inertial regime due 
to aggregate-induced rigidity and reduced water flow, 
and a slowing down
in the coalescence rate (deceleration) when
surfactant concentration increases,
while at the same time
we also confirm the existence of
an initial thermal, and a power-law, inertial,
regime of the bridge growth dynamics in
both the pure and the surfactant-laden droplets.
Thus, we unveil the key mechanisms in one of the 
fundamental topological processes of 
liquid droplets containing surfactant,
which is crucial in relevant technologies. 
\end{abstract}

\maketitle

\section{Introduction}
Droplet coalescence plays an important role 
in many natural phenomena, 
for example, determining the size distribution
of droplet rains \cite{Bowen1950,Berry1974}, the dynamics of multiphase flows \cite{Campana2004,Lu2012}, 
and, also, in technological applications,
such as inkjet printing \cite{Singh2010} or coating applications \cite{Frohn2000}.
The coalescence process depends on the interplay
between viscous and inertial forces and surface tension, 
with the minimization of the latter driving this process.
Experiments, theories, and simulations 
of the coalescence of droplets 
without additives have provided great insight into its 
mechanisms \org{\cite{Paulsen2014,yoon2007coalescence,khodabocus2018scaling,perumanath2019droplet,eggers1999coalescence,aarts2005hydrodynamics,Sprittles2012,dudek2020,Rahman2019,Berry2017,Somwanshi2018,Kirar2020,Bayani2018,Brik2021,Anthony2020,Kern2022,Heinen2022,Geri2017,Abouelsoud2021}}, 
but much less is known in the case of surfactant-laden droplets\org{\cite{Politova2017,Weheliye2017,Soligo2019,Dong2017,Dong2019,Kovalchuk2019,Botti2022,Amores2021,Kasmaee2018,nowak2017bulk,nowak2016effect,jaensson2018tensiometry,narayan2020zooming,ivanov1999flocculation,tcholakova2004role,langevin2019coalescence,Lu2012,Velev1993,Suja2018}}
\blu{or \org{droplets with other additives}
\cite{Dekker2022,Calvo2019,Arbabi2023,Varma2021,Sivasankar2021,Sivasankar2022,Otazo2019,Vannozzi2019}}, 
despite their relevance in many areas, 
such as cloud formation \cite{kovetz1969effect}, 
microfluidics \cite{feng2015advances}, 
coating technologies \cite{ristenpart2006coalescence}, 
and water treatment during crude oil and natural gas 
separation \cite{dudek2019microfluidic}.  
Based on high-speed imaging and particle image velocimetry
technology,
experimental studies have investigated the coalescence of 
surfactant-laden droplets, 
mainly providing macroscopic descriptions of the coalescence process
\cite{nowak2016effect, narayan2020zooming, nowak2017bulk, duchemin2003inviscid, leal2004flow,Chinaud2016}. 
However, the initial fast stages of
the coalescence process are impossible to observe
in experiments due to device limitations.\cite{Bayani2018} 
Moreover, conventional hydrodynamic models 
are only applicable in the later stages of
coalescence \cite{yeo2003film,hu2000drop,mansouri2014numerical},
while the singularity at the initial 
contact point of the coalescing 
droplets continues to pose challenges 
for numerical modelling despite 
progress in this area \org{\cite{eggers1999coalescence,duchemin2003inviscid,Sprittles2012,Heinen2022}}. 
To address the latter issue, for example,
continuum modeling may consider either the formation of
a single body of fluid by an instant appearance of a liquid
bridge that smoothly connects the two droplets and then 
evolves as a single body due to capillary forces, 
or a section of the free surface trapped between the bulk
phases that gradually disappears \cite{Sprittles2012}. 
In the case of systems with surfactant,
continuum simulation has suggested that an uneven
contraction of the interface due to a nonuniform
distribution of accumulating surfactant at the meniscus 
bridge that connects the droplets is an important factor
that modulates the surface tension, which,
in turn, drives the coalescence process \cite{Lu2012}. 
Still, numerical simulation is unable to analyze 
the mechanism of coalescence after the drops come 
into contact.
Recent molecular-level simulations have 
clarified important aspects, such as 
the role of thermal capillary waves 
at the surface of water droplets
\cite{perumanath2019droplet},
but the effect of surfactant on the physics involved in the
coalescence has remained overwhelmingly unexplored. 
We know surfactant effects must be large since they 
greatly change the surface tension, so the research 
reported here set out to clarify its role in the 
coalescence dynamics and other characteristics.

In this study, we report on large-scale 
MD simulations based
on a high-fidelity coarse-grained (CG) force-field \cite{theodorakis2015modelling,Theodorakis2015Langmuir,theodorakis2019molecular,Theodorakis2014,Theodorakis2019},
which allows for the faithful simulation 
of surfactant in water.
With these we uncover the 
mass transport mechanism of surfactant during
coalescence, elucidate the dynamics of the bridge growth
process, resolve the flow, 
and analyse how the above depend on
surfactant distribution.
We find an unexpected lack of multiple thermally
excited precursor bridges, 
attenuated collective flow after contact, 
formation of new aggregates inside the bridge
from surfactant previously at the droplets' surface,
and a slowing down in the inertial regime 
as surfactant concentration increases.
In the following, we provide some background
information in Sec.~\ref{background}. Then,
we present our simulation model and methods
in Sec.~\ref{model} and our results and relevant
discussion in 
Sec.~\ref{results}. Finally, we draw our conclusions
and suggest possible directions for future
work in Sec.~\ref{conclusions}.

\begin{figure}[bt!]
\includegraphics[width=0.6\textwidth]{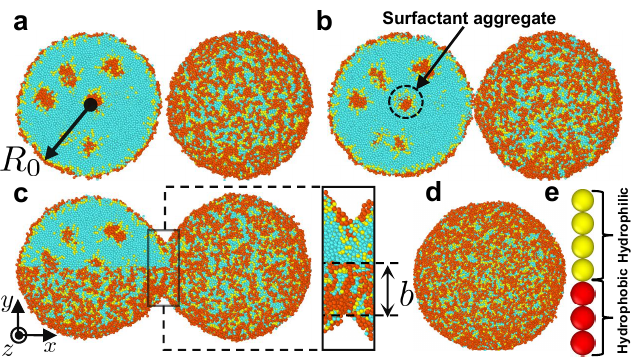}
\caption{\label{fig:1}
Stages of coalescence of spherical
surfactant-laden droplets with equal size and 
surfactant concentration ($3.2$~CAC). 
a) Initial configuration;
b) Beginning of the bridge formation;
c) Bridge growth with a magnified view of the bridge region. 
$b$ is the radius of the bridge; 
d) Final equilibrium configuration after reshaping;
e) Coarse-grained representation of a C10E4 surfactant molecule.
The surfactant's hydrophobic beads are in red, hydrophilic ones
in yellow. Each cyan bead represents two water molecules.
External or cross-section views
are shown to highlight the bulk, surface, and bridge
structure of the droplets. Surrounding water vapor
is omitted for the sake of clarity. 
The snapshots of the system were obtained using 
Ovito software \cite{Stukowski2010}.}
\end{figure}

\section{Background}
\label{background}
Droplet coalescence takes place in three different stages,
namely, the
droplet approach, when the two droplets are positioned 
close enough to `feel' intermolecular forces
(Fig.~\ref{fig:1}a), the bridge growth-stage 
(Figs~\ref{fig:1}b and \ref{fig:1}c) \cite{jin2017coalescence}, 
and the final reshaping stage towards the equilibrium 
spherical droplet (Fig.~\ref{fig:1}d).
In the case of droplets without surfactant,
the growth dynamics of the bridge
has been investigated and 
in general, two different regimes have been assumed 
from the perspective of fluid 
dynamics \cite{duchemin2003inviscid,eggers1999coalescence}:
an initial viscous regime dominated by macroscopic flows 
that pull the droplets together, and a subsequent inertial
regime, which involves the propagation of local deformations
with higher Reynolds number excited near the bridge as it grows.

Even in the case without
surfactant, the bridge growth dynamics
has been under intense debate.
In the viscous regime (VR), 
a linear scaling in time $b\propto t$ has been 
suggested for the bridge radius, $b$, 
as well as logarithmic corrections $t\ln t$
\cite{duchemin2003inviscid,eggers1999coalescence},
while a scaling $b\propto \sqrt{t}$ has been proposed for
the inertial regime (IR)
\cite{duchemin2003inviscid,eggers1999coalescence}.
However, others  have suggested scaling regimes that
depend on the ratio of characteristic scales to the viscous
length scale, including an additional
inertially limited viscous (ILV) regime, 
\cite{paulsen2012inexorable,paulsen2013approach}
\blu{which\org{, according to numerical simulations,} is only realized when the coalescing drops are
initially separated by a finite distance
\cite{Anthony2020}}\org{. Another idea put forward has been}
the characterization of the viscous--inertia-regime
transition  via a 
modified Ohnesorge number in the case of
immiscible droplets \cite{Xu2022}.
Despite
the advent of modern experimental techniques, 
such as electrical 
measurements with resolution of a
few micrometers \cite{Paulsen2011},
the bridge growth dynamics at the early 
stages still remains challenging
for experimental studies. Instead, molecular dynamics (MD) 
simulation of an all-atom model for water droplets
has provided insight into this initial stage of coalescence,
suggesting the formation of multiple precursor bridges at 
the pinch point, due to thermal
capillary waves at the droplet surfaces \cite{perumanath2019droplet}. 
These multiple bridges 
expand linearly in time, due to
collective molecular jumps at the droplets' interface, 
and the transition to the classical 
hydrodynamics regime only takes place 
when the bridge radius becomes
larger than a thermal length,
$l_T  \approx {\left ( k_BT/\gamma \right) }^{1/4} R^{/1/2}$, 
assuming that fluctuations on one droplet are not
affected by the other
and in the absence of instabilities \cite{perumanath2019droplet}.
$l_T$ describes the typical width of the contact points
at droplet's interface at the initial stage of
coalescence, $k_B$ is Boltzmann's constant, $T$ the
temperature, $\gamma$ the liquid--vapor (LV) 
surface tension, 
and $R$ the radius of the droplet.
Since \blu{$l_T$} depends on surface tension, it
is expected to grow  with surfactant
concentration as $\gamma$ decreases, saturating to 
a value, $l_s$, above the critical aggregation
concentration (CAC) as $\gamma$ reaches
its plateau value.

In the presence of surfactant there are many unknowns. 
Several studies have suggested that its presence
would actually delay the coalescence process, due to the reduction 
of the surface tension \cite{jaensson2018tensiometry,leal2004flow},
while smaller droplets tend to show much faster equilibration
of surfactant interfacial coverage \cite{jin2004surfactant,narayan2020zooming}.
Moreover, it has been suggested that physical 
regimes could also depend on the diffusion and adsorption time scales of the surfactant, and their 
dependence on the surfactant concentration and the droplet size \cite{jin2004surfactant}. 
In addition, it has been pointed out that 
surfactant alters the properties of the droplets
particularly in the bridge area \cite{ivanov1999flocculation}. 
For example, 
hydrodynamic instabilities, such as dimples, 
have been observed for concentrations larger
than CAC \cite{langevin2019coalescence}, but 
surfactant might actually have a more global effect by
affecting the overall size of the droplets \cite{tcholakova2004role}.
Certain experiments have also highlighted the role of
Marangoni flow that leads to local capillary pressure changes, 
which in turn affect the coalescence kinetics and 
result in a delay of the process \cite{nowak2016effect,Lu2012}. 
Despite these efforts,
the mass transport mechanism of surfactants, 
the resulting dynamics and structure
of the bridge and other early time 
effects are not well understood.
Molecular simulations allow for tracking the
individual molecules,
which goes beyond the reach of any continuum simulation or
real experiment and is therefore crucial for unravelling
the mass transport mechanism of surfactant.
At present, the early time phenomena
that are pivotal for the onset of
coalescence can only be investigated in adequate detail by 
molecular-scale simulation.

\section{Model and Methodology}
\label{model}
Our investigation covers all stages of
coalescence for droplets of equal size 
and surfactant concentration.
We have  considered different surfactants, such as
C10E8 and C10E4 \cite{Theodorakis2015Langmuir}, 
and a range
of surfactant concentrations below/above CAC.
The interactions between
components of the system are obtained by the
Mie-$\gamma$
Statistical Associating Fluid Theory (SAFT Mie-$\gamma$) \cite{Muller2014,Lobanova2015,Lafitte2013,Avendano2011,Avendano2013}.
The MD simulations were carried out in the
canonical ensemble
using LAMMPS software \cite{Plimpton1995,LAMMPS}. 
After equilibration of each individual droplet,
the droplets were placed next to each other
for initiating their coalescence as 
illustrated in Fig.~\ref{fig:1}a. 

The force field
has been validated for water--surfactant systems
with particular focus on accurately reproducing 
the most relevant properties of the system,
such as surface tension and phase behavior
\cite{theodorakis2015modelling, Theodorakis2015Langmuir, Theodorakis2019, theodorakis2019molecular, Theodorakis2014, lobanova2014development, lobanova2016saft, morgado2016saft}.
Interactions between the various types of CG beads
are described
via \org{the} Mie potential, 
which is mathematically expressed as

\begin{equation}
\label{equation_mie}
    \blu{U(r_{ij})} = C\epsilon_{\rm ij} \left[ \left({\frac{\sigma_{\rm ij}}{r_{\rm ij}}}\right)^{\lambda_{\rm ij}^{\rm r}} - \left({\frac{\sigma_{\rm ij}}{r_{\rm ij}}}\right)^{\lambda_{\rm ij}^{\rm a}}\right], 
    \; {\rm for} \; r_{\rm ij} \leq r_{\rm c},\\
    \end{equation}
    where
\begin{equation*}
    C = \left(\frac{\lambda_{\rm ij}^{\rm r}}{\lambda_{\rm ij}^{\rm r} - \lambda_{\rm ij}^{\rm a}}\right){\left( \frac{\lambda_{\rm ij}^{\rm r}}{\lambda_{\rm ij}^{\rm a}}\right)}^{\frac{\lambda_{\rm ij}^{\rm a}}{\lambda_{\rm ij}^{\rm r} - \lambda_{\rm ij}^{\rm a}}}.
\end{equation*}
i and j are the bead types, $\sigma_{\rm ij}$ indicates the 
effective bead size
and $\epsilon_{\rm ij}$ is the interaction strength
between beads i and j. 
$\lambda_{\rm ij}^a=6$ and $\lambda_{\rm ij}^r$ are 
Mie potential parameters,
while $r_{\rm ij}$ is the distance between two CG beads.
A universal cutoff for all nonbonded interactions is set 
to $r_c = 4.583$~$\sigma$.
Units are chosen for the length, 
$\sigma$, energy, $\epsilon$, mass, $m$, and
time, $\tau$, which in real units would roughly correspond to:
$\sigma = 0.43635$~nm, $\epsilon / k_B = 492$~K, $m=44.0521$~amu
and $\tau = \sigma{(m/{\epsilon})}^{0.5} = 1.4062$~ps. 
All simulations are carried out in the NVT ensemble by using the
Nos\'e--Hoover thermostat as implemented in the LAMMPS
package \cite{Plimpton1995,LAMMPS} with an integration time-step 
$\delta t = 0.005$~$\tau$. Moreover, simulations took place
at room temperature, therefore,
$k_BT/\epsilon = 0.6057$, 
which corresponds to $T=25$~$^\circ$C.

Surfactants of type C$n$E$m$ are considered, 
such as C10E8 and C10E4. 
A hydrophobic alkane CG `C' bead 
represents a $\rm -CH_2-CH_2-CH_2-$ 
group of atoms, while a hydrophilic CG `EO' bead represents an
oxyethylene group $\rm -CH_2-O-CH_2$. Finally, a water CG `W'
bead corresponds to two water molecules.
In Table~\ref{tab:table1}, the nonbonded 
interactions between the different CG beads are listed, 
while the mass of each bead is reported in 
Table~\ref{tab:table2}.

Bonded interactions are taken into account via a harmonic
potential, \textit{i.e.}, 
\begin{equation}
\label{equation_bonded1}
    V(r_{\rm ij}) = 0.5k(r_{\rm ij}-\sigma_{\rm ij})^2
\end{equation}
where  $k = 295.33$~$\epsilon/\sigma^2$.
Moreover, EO  beads experience a harmonic angle potential:
\begin{equation}
\label{equation_bonded2}
    V_\theta(\theta_{\rm ijk}) = 0.5k_\theta(\theta_{\rm ijk}- \theta_0)^2 
\end{equation}
where $\theta_{\rm ijk}$ is the angle between consecutive
beads i, j and k.
$k_\theta = 4.32$~$\epsilon/$rad$^2$, while $\theta_0 = 2.75$~rad
is the equilibrium angle. 
Further discussion on the model can be found in previous 
studies \cite{Theodorakis2015Langmuir,theodorakis2015modelling}.

\begin{table}[bt!]
\caption{\label{tab:table1}
Mie-potential interaction parameters between CG beads. $\lambda_{\rm ij}^{\rm a} = 6$ for all cases.
}
\begin{ruledtabular}
\begin{tabular}{lccr}
\textrm{i--j}&
\textrm{$\sigma_{\rm ij}~(\sigma)$}&\textrm{$\epsilon_{\rm ij}~(\epsilon/k_B)$}&\textrm{$\lambda_{\rm ij}^{\rm r}$} \\
\colrule
W--W & 0.8584 & 0.8129 & 8.00\\
W--C & 0.9292 & 0.5081 & 10.75 \\
W--EO & 0.8946 & 0.9756 & 11.94 \\ 
C--C & 1.0000 & 0.7000 & 15.00 \\
C--EO & 0.9653 & 0.7154 & 16.86\\
EO--EO & 0.9307 & 0.8067 & 19.00
\end{tabular}
\end{ruledtabular}
\end{table}

\begin{table}[bt!]
\caption{\label{tab:table2}%
Mass of CG beads.
}
\begin{ruledtabular}
\begin{tabular}{cc}
\textrm{Bead type}&
\textrm{Mass~(m)} \\
\colrule
W & 0.8179 \\
C & 0.9552  \\
EO & 1.0000
\end{tabular}
\end{ruledtabular}
\end{table}

To prepare the initial configuration of each system,
individual
droplets were first equilibrated in the NVT ensemble. 
The total number of beads in the simulations 
was $10^5$ per initial droplet, 
with approximately 5\% evaporation into the gas. 
Droplet diameters were $\sim53~\sigma$, which is about 23~nm.
Careful consideration was given \org{during the preparation} not only
to observing the energy of the system, but,
also, making sure that the distribution of clusters
has reached a dynamic equilibrium and that each
cluster was able to diffuse \org{a} 
distance many
times \org{its size.} 
After equilibration, the system
size (volume of the simulation box) was doubled
and droplets were placed next to each other 
in such a way to avoid interaction between mirror images
of the droplets that could potentially occur due to the presence of 
periodic boundary conditions in all directions, 
if one was not careful. 
In this way, the same thermodynamic conditions 
for the system were approximately guaranteed in the system
of a single droplet and in the systems of
two droplets used for coalescence. 
Figure~\ref{fig:1}a illustrates 
a typical initial configuration of the system.
Only the liquid state (droplets) is shown, which
is identified by a 
cluster analysis \cite{Theodorakis2010,Theodorakis2011}, while
surrounding vapour has been removed for the sake 
of clarity.
Finally, for our droplets, we have considered a range
of different surfactant concentrations below and above
the CAC and up to about 6.1 CAC.
This covers the whole range of 
concentrations  
relevant for the mass transport and other phenomena
discussed here.

To quantify the mass transport of surfactant, 
first a grid with mesh size of $2~\sigma$
is defined and surfactant and water
particles are assigned to each grid cell.
The grid size is chosen to guarantee
adequate accuracy in the position of the grid
cell while avoiding excessive technical randomness
due to having a mesh finer than the size of single beads.
Then, based on the density,
one can identify the grid cells that belong to the
droplets surface or the bulk. By following the
grid locations of the surfactant beads,
we are able to track the transport of surfactant 
between the different parts of the droplets. 
The central bead in a molecule determines
whether it is counted as bulk or bridge,
whereas if any bead of a molecule enters
a surface grid cell the
molecule is counted as being on the surface. 

\begin{figure}[bt!]
\includegraphics[width=0.6\columnwidth]{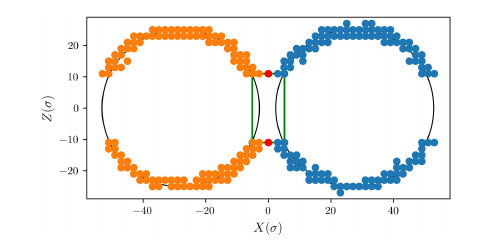}
\caption{\label{fig:2} Specifying the bridge (green rectangle).
Orange and 
blue points are surface grids on left and right droplets, respectively.  
Red points are the location of highest and lowest beads on bridge.
The solid black line is a best fit to the surface grid positions
following Ref.~\citen{Kanatani2011}.
}
\end{figure}

To track the bridge growth, we need to define
the bridge region. In our case, this
is a slab whose width in the $X$ direction is
recalculated at each snapshot. 
The left and right limits of the slab are
determined by analyzing the grid points
on the $X-Z$ plane after droplets have been
aligned with the coordinate system as
shown in Fig.~\ref{fig:2}.
We fit a circle
around each droplet
and note the surface grid  positions at the central $X=0$ position, 
shown by the red points in Fig.~\ref{fig:2}.
Horizontal lines are drawn in the $X$ direction
passing through these red points to touch the fitted circles,
thus defining the rectangle in green. 
The vertical sides of the rectangle give the limits 
of the bridge slab in the $X$ direction and its width.
All molecules with centers having $X$ coordinates inside these 
limits are labelled as belonging to the bridge in a given snapshot.

On the other hand, the bridge radius, $b$,
(Fig.~\ref{fig:1})
is calculated using the distances between extrema
of the positions of the beads belonging to the grids 
located at $X=0$, \textit{i.e.} this distance is first 
calculated separately for the $Z$ coordinate to
give a distance $2b_Z$, and then for the $Y$ 
coordinate to give $2b_Y$. The final bridge 
radius estimate is then given by $b=(b_{Z}+b_{Y})/2$.

\begin{figure*}[bt!]
\includegraphics[width=\textwidth]{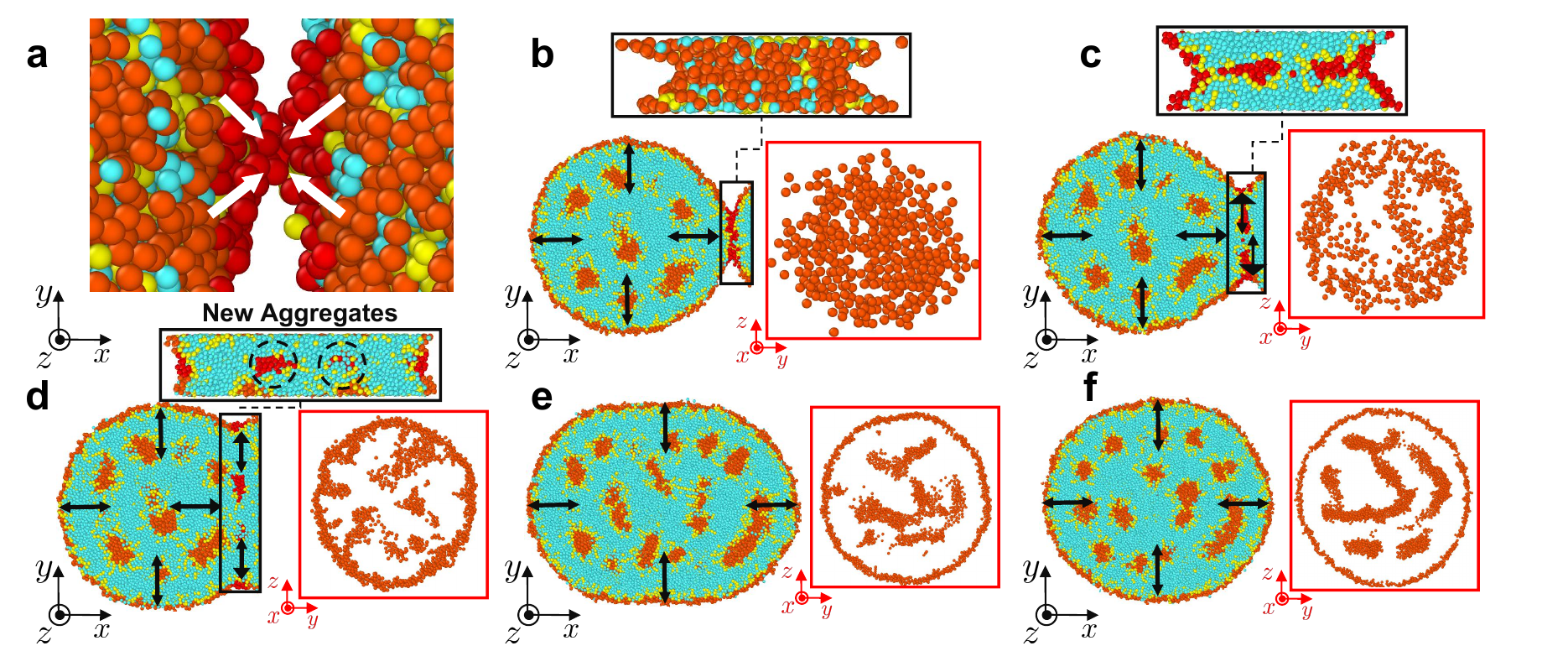}
\caption{\label{fig:3} Mass transport mechanism of 
surfactant (C10E4\, \blu{$4.7$ CAC}) during
the coalescence process.
(a) Droplet pinching (precursor bridge) 
taking place through the aggregation
of surfactant at the first contact point 
of the droplets ($t-t_c=6.25~\tau$). \blu{\org{Here} $t_c$ is the time of first contact;} 
(b) Main surfactant transfer processes during the initial
stage of coalescence as indicated by arrows on a droplet
cross-section \org{in} 
the $x-y$ plane ($t-t_c=32.5~\tau$). 
A larger arrow end indicates the dominant direction of 
surfactant transport between the different regions 
in the droplet.
Magnified views of the bridge and 
its cross section on the $y-z$ plane
(only hydrophobic beads) are also shown \org{above}. At this stage, 
the bridge is dominated by the presence of surfactant 
molecules.
(c), (d), (e), and 
(f) at times $t-t_c=76.25~\tau, t-t_c=233.75~\tau, t-t_c= 517.50~\tau, 
t-t_c=733.75~\tau$, respectively, show evolution in
the inertial regime. The snapshots of the system were obtained using 
Ovito software \cite{Stukowski2010}.
}
\end{figure*}

\section{Results and Discussion}
\label{results}
The mass transport mechanism of surfactant
molecules during coalescence is fundamental
to understanding the role of surfactant
in the dynamics of this process at all stages.
\blu{Surfactant mass transfer mechanisms have been
investigated in various processes, for example,
superspreading,\cite{Theodorakis2015Langmuir}
emulsion films,\cite{Velev1993} and foam 
stabilization in lubricating oils.\cite{Suja2018}
For example, in the case of emulsion films, 
a fascinating cyclic phenomenon has been observed
where new dimples sequentially form with the surfactant
redistribution driving this process through coupling
to interfacial an hydrodynamic motion inside the
films.\cite{Velev1993}.}
\org{In our system,} coalescence starts with the 
formation of the contact point
(Fig.~\ref{fig:3}a), where
hydrophobic beads from the two droplets
actively move to aggregate due
to the favorable attractive interaction. 
In the case of surfactant-laden droplets, 
we have not observed the formation of
multiple contact points (bridge precursors)
for any of the systems,
unlike what has been 
seen in pure water
droplets \cite{perumanath2019droplet}.
\org{In fact,} water molecules
do not participate at this \org{earliest} stage in the bridge 
formation.
The bridge growth process continues with the formation 
of a thin layer of surfactant between
the droplets (Fig.~\ref{fig:3}b), 
whose origin is mostly from the initial surface coverage.
To unveil these processes, we have monitored the
transport of surfactant between different parts of the droplets,
\textit{i.e.} the interior, the bridge,
and their surfaces, which sums up to
36 possible surfactant transport processes.
The Supplemental Material (Table S3) provides the numbers
for the probabilities of surfactant
remaining at a certain place or moving
to different parts of the droplets for all
cases considered in our study.
At this stage, the still small radius, $b$, of the bridge 
permits a high supply of surfactant
at the contact surface (Fig.~\ref{fig:3}b), which
is central to the coalescence of the droplets.
However, as the bridge further grows,
the surfactant from the initial contact and inflow 
to the bridge perimeter is not enough
to fully supply the interior
of the bridge with surfactant. 
The perimeter of the bridge grows proportionally to $b$,
while its area (cross-section) increases with $b^2$. 
Therefore,
the concentration of surfactant
in the bridge, initially very high,
reduces proportionally to $1/b$ as the bridge grows. 
Moreover, tracking the molecules shows
that, as the bridge forms, 
less molecules end up in the bridge bulk 
than were on the approaching surfaces prior to contact. 
Surfactant transport towards the surface 
is favourable energetically and only surfactant 
that cannot escape to the exterior (surface)
remains trapped in the interior of the
bridge region. 
As a result, the engulfed surfactant forms
separated aggregates within the bridge,
especially for the cases above CAC
(Fig.~\ref{fig:3}d).
These aggregates are characteristic of the bridge growth
at later stages (Figs~\ref{fig:3}c--e),
and, as we will see later by the analysis of
the bridge growth, surfactant from the bulk can join
the aggregates that formed at the bridge as it grows.

\begin{figure}[bt!]
\includegraphics[width=0.5\columnwidth]{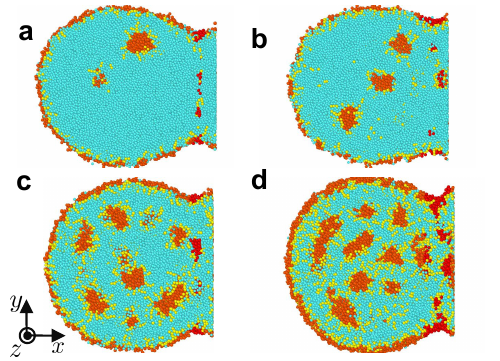}
\caption{\label{fig:4} Droplet interiors 
in the
inertial regime showing the presence of new aggregates emerging
during the coalescence process. 
Cross-sections are shown at times corresponding
to Fig.~\ref{fig:3}d for
surfactant in different concentrations of C10E4 above the CAC: 
(a) 1.6~CAC, (b) 3.2~CAC, (c) 4.7~CAC and (d) 6.1~CAC.
The snapshot of the systems were obtained using 
Ovito software \cite{Stukowski2010}.
}
\end{figure}

The relevant surfactant transport processes
during the bridge growth that we have identified 
are the engulfment of 
surfactant from the contact surface of the droplets into the
interior of the bridge (
Tables S1 and S2 in the
Supplemental Material give details), which increases with surfactant
concentration, and
to a smaller extent the transfer of surfactant 
in the bulk towards the bridge (Figs~\ref{fig:3}c, d).
Coalescence is mainly
affected by the transfer of surfactant in the region
close to the bridge from the interior to the
surfaces, while, in the other parts
of the droplet, surfactant is rather
in dynamic equilibrium and does
not affect the coalescence process. 
After the bridge fully develops (Fig.~\ref{fig:3}e),
a dynamic equilibrium of surfactant extends
throughout and no dominant directions
of adsorption/desorption processes remain, but only
a slight surfactant transport from the surface
towards the bulk as the surface area of the droplet
becomes smaller.
At this final stage, the droplet will
reach its final spherical shape (Fig.~\ref{fig:3}f), 
driven by the surface tension. 
We have also verified that the new aggregates emerging
during the coalescence process consist of surfactant
that was previously on the contact area (surfaces) 
between the two merging droplets.
The latter observations are valid 
throughout a range of different
concentrations and surfactants below and above the CAC. 
Data for other
concentrations and surfactants 
than in 
Fig.~\ref{fig:3}
are reported in the Supplemental Material, and 
show the same mechanism, while snapshots of the
aggregate formation in the inertial regime are
presented in Fig.~\ref{fig:4}.

\begin{figure}[bt!]
\includegraphics[width=0.6\columnwidth]{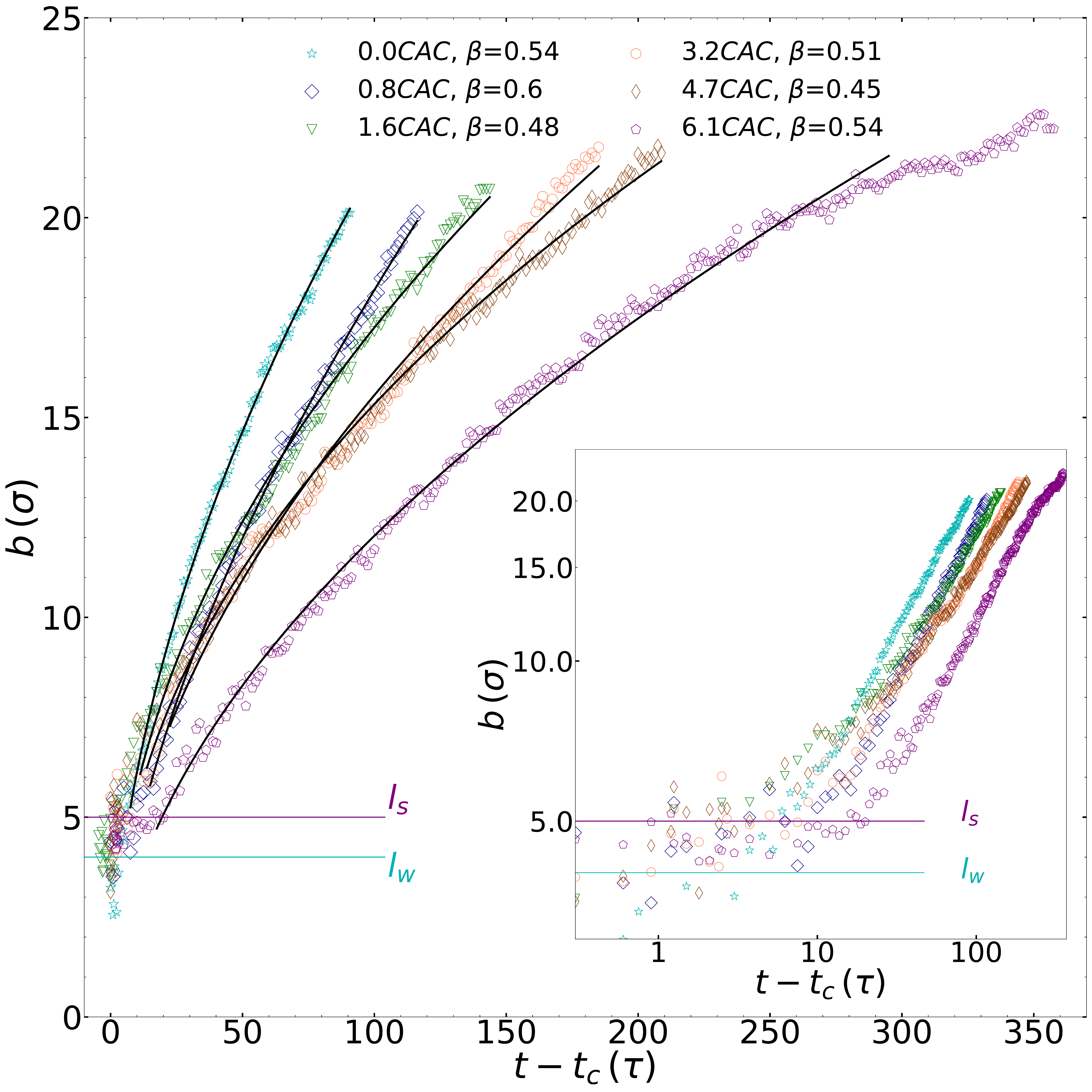}
\caption{\label{fig:5} Bridge growth dynamics
($b$, radius of the bridge) \textit{vs.} time, $t$, for 
droplets with different surfactant concentrations (C10E4),
as indicated in the legend. 
CAC $\approx7.5$~wt\%.
Power law fits are also shown, 
tentatively identifying  the inertial
(IR, $b=b_0 t^{\beta}$) regimes.
The inset highlights the power law scaling in the 
inertial regime and the initial TR regime.
$l_w$ is the thermal length for pure water droplets
and $l_s$ for surfactant-laden droplets above CAC
according to Ref.~\citen{perumanath2019droplet}.
Data for C10E8 and average growth rates
are 
provided in the Supplemental Material.
}
\end{figure}

To identify the various regimes and 
better understand the bridge growth dynamics, we have 
measured the bridge radius, $b$, over time for droplets with
different surfactant concentrations 
(Fig.~\ref{fig:5} here for C10E4 and Fig.~S1 for C10E8 in the
Supplemental Material). 
The regimes that follow after the initial bridge formation
can be in principle identified by
the bridge radius scaling.
The inertial scaling with power law $b\sim\sqrt{t}$ 
is generally \org{most} 
conspicuous,
although we see an apparent
changeover from an initial thermal regime (TR) \org{with little bridge growth}
to the IR power law  (see fits in Fig.~\ref{fig:5} 
and Fig.~S1 of the Supplemental Material).
Moreover, in Fig.~\ref{fig:5}, the values of the thermal
lengths are marked with the horizontal lines for the
cases of pure
water and surfactant-laden droplets (above CAC) according to
previous MD predictions \cite{perumanath2019droplet}.
These values \org{are of the same order as the TR regime bridge size that} 
we observe in our data, and express the range of 
\org{the} thermal length scale \org{above} 
which a persistent
increase of the bridge radius, $b$, takes place.

Our findings also indicate that the
growth speed of the bridge
decreases as a function of surfactant concentration
in both regimes. 
Tracking the simulation trajectories we observe 
that the surfactant aggregates that are present
in the bulk can slow the liquid flow and obstruct the
strong water--water interactions. Upon
a significant increase of surfactant
concentration far above CAC,
aggregates merge in the bulk
leading to an increased 
rigidity of the droplet. 
This then hinders the coalescence
process by slowing down the rearrangement of 
the droplet towards its equilibrium spherical shape. 
This is explained by the interactions of water 
and hydrophobic beads (see Supplemental Material), 
which indicate a larger W--W (W: water beads) than \mbox{C--C}
(C: hydrophobic surfactant beads) attraction
and a strongly unfavorable (less attractive) C--W 
interaction in
comparison to the interactions of all other components.
The average 
bridge growth velocity, which includes both the
TR and IR regimes together as an overview of the overall
speed of growth is reported for each surfactant for
a range of concentrations in Table~\ref{tab:table3}.
It is calculated over the time interval between the moment
that the link between the droplets is
established\org{, $t_c$,} at the beginning of the coalescence until
the point at which the bridge radius is equal to the 
radius of the droplets in the $y$ direction 
(for example, see Fig.~\ref{fig:3}e).
As surfactant concentration increases,
the bridge growth process slows down in comparison with 
the simulated case of the pure water droplets.
These data also show a slightly faster bridge 
growth in the case of the C10E8 surfactant 
(see Fig.~S1 in Supplemental Information). 

\begin{table}[bt!]
\caption{\label{tab:table3}%
Average velocity of bridge growth in units $\sigma/\tau$
\footnote{For pure water droplets in the viscous regime (result from simulation): \org{0.3675} $\sigma/\tau$.}} 
\begin{ruledtabular}
\begin{tabular}{lcccccccc}
Concentration (CAC) & 0.8 & 1.6 & 3.2 & 4.7 & 6.1 \\
\colrule
C10E4 & \org{0.2849} & \org{0.2204} & \org{0.1878} & \org{0.1605} & \org{0.1047}\\
C10E8 & \org{0.2794} & \org{0.2319} & \org{0.1871} & \org{0.1530}& \org{0.1115}\\
CAC = 7.5~wt\%
\end{tabular}
\end{ruledtabular}
\end{table}

\begin{figure}[bt!]
\includegraphics[width=0.5\columnwidth]{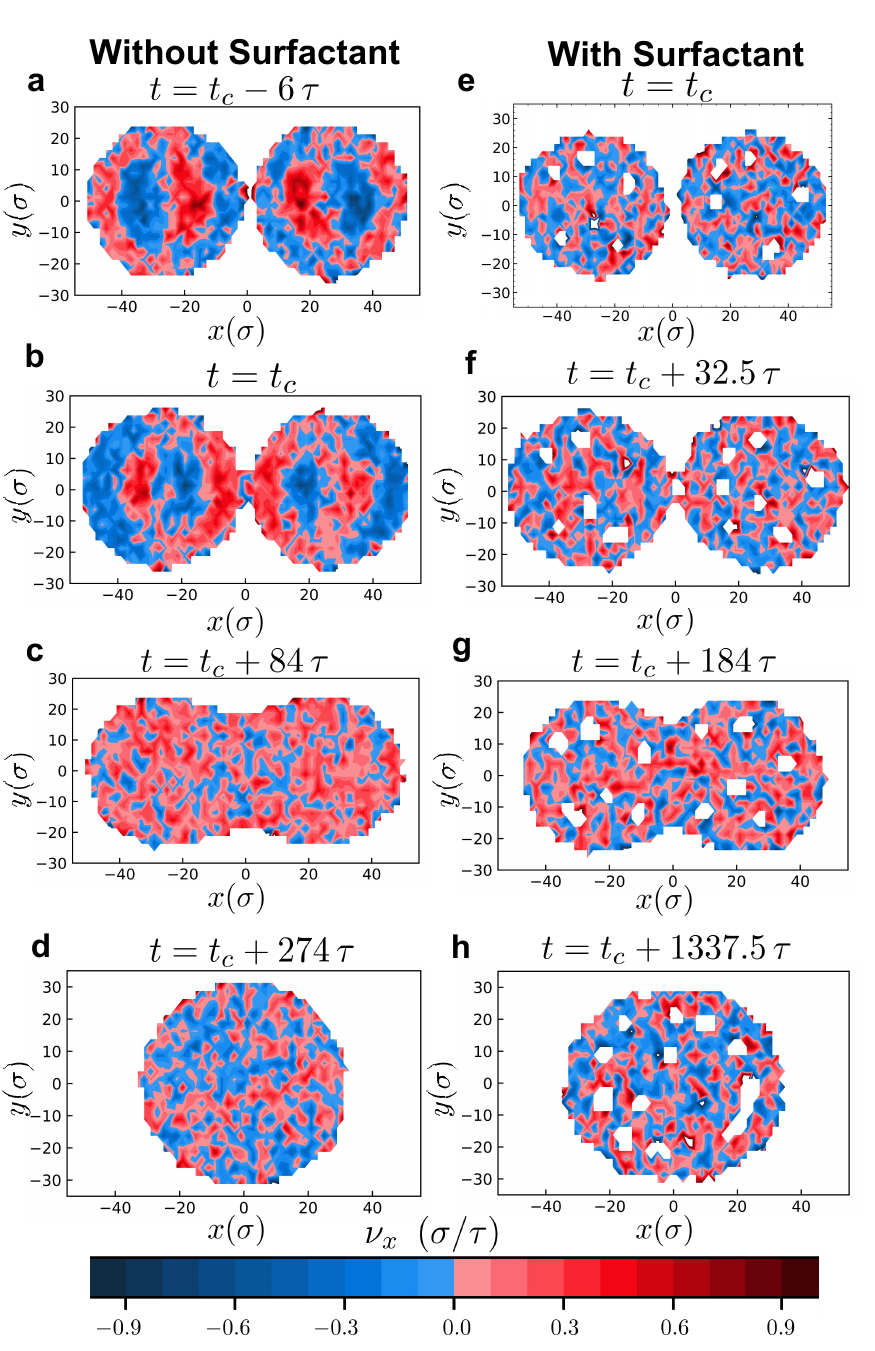}
\caption{\label{fig:6} Flow field of water 
($x$ velocity component, $\varv_{x}$),
in cross-sections of droplets without (a--d) 
and with (e--h) surfactant (C10E4) at
4.7~CAC concentration,
at different stages of coalescence. Side by side times
correspond to similar stages of the coalescence
process. 
Red reflects the intensity of flow (only water)
motion towards the bridge, blue away from the bridge. \blu{Time labels based on contact time ($t_c$) are added.} \org{Note that} white space between the water areas 
(\org{e.g. in} 
the bridge) \org{includes} 
surfactant aggregates \org{and surfactant on the surface. This can cause an illusion of multiple contact points such as in panel f, which are in fact surrounded by surfactant forming an overall broad bridge.}}
\end{figure}

Furthermore, the flow field of water molecules during coalescence
exhibits differences between droplets with and without
surfactant. In Fig.~\ref{fig:6}, the colour code indicates
flow towards the bridge (red) and away from the bridge 
(blue). In the case of the water droplets without
surfactant, the formation of the bridge at the very initial
stages  is accompanied by fluctuations of
internal collective flow in
the direction of the coalescence axis ($x$ direction),
which encompass the entire droplets (Figs \ref{fig:6}a, b). 
This is due to the capillary waves produced by the 
energy release from the initial rupture of the surface 
when the droplets first touch.\cite{Dekker2022} The waves
propagate and result
in perturbations in the overall shape of the 
droplets and the flow patterns
illustrated in Figs~\ref{fig:6}a and b. 
These flow patterns disappear as the bridge 
grows further and a robust contact between the
two droplets establishes beyond the thermal
regime. Moreover, an overall flow
towards the bridge as the droplets further coalesce is
observed (notice the dominance of red in Fig.~\ref{fig:6}c),
while at the final equilibrium only random thermal fluid flow
patterns are seen (Fig.~\ref{fig:6}d).
We have not noticed
any statistically significant flow patterns
or Marangoni flow \cite{Amores2021} (\textit{e.g.\/} in the case of
droplets with surfactant) towards any of the
other directions
(\textit{e.g.\/} radial).

\begin{figure}[bt!]
\includegraphics[width=0.5\columnwidth]{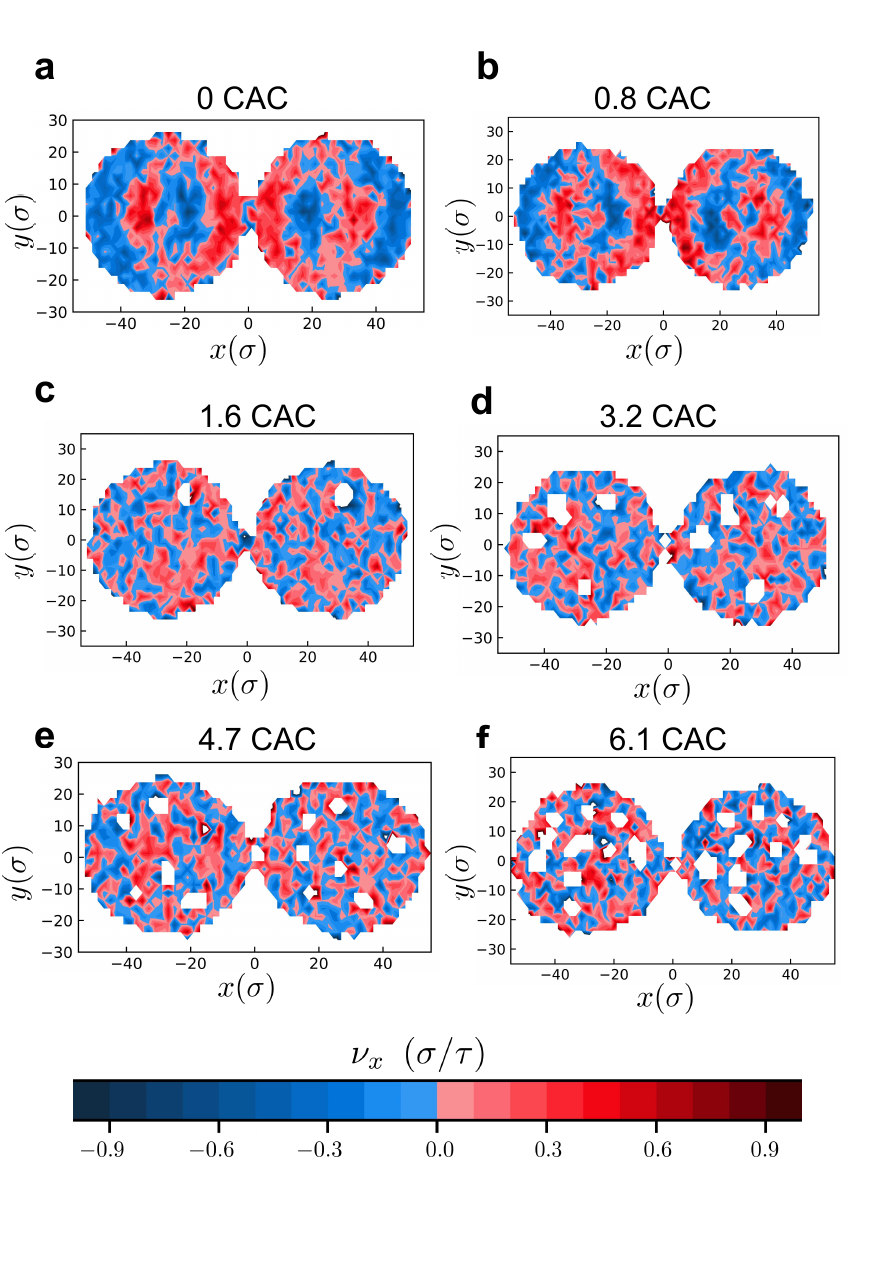}
\vspace*{-1.5cm}
\caption{\label{fig:7} Water flow pattern for different 
concentrations (a--f, 0--6.1~CAC) for C10E4 at the initial pinching stage. Above CAC (c--f),  there 
is no observable pattern. Red color reflects the
intensity of flow towards the bridge, while blue 
indicates the intensity of flow away from the bridge. \org{Like in Fig.~\ref{fig:6},} empty space within/between the \org{water in the} droplets
(\org{e.g.} 
in the bridge) \org{includes} 
surfactant\org{, 
and the contact point of the droplet surface is always single}. 
}
\end{figure}

As surfactant is added to the droplets the early time collective flow
patterns of Figs~\ref{fig:6}a and \ref{fig:6}b gradually
disappear,
especially for all concentrations above
the CAC (Fig.~\ref{fig:7}). 
Compare, for example,
the flow patterns in Fig.~\ref{fig:6}b and \ref{fig:6}f, 
for the same bridge size.
The suppression of the collective flow
occurs through two routes:
First, the surfactant at the surface reduces the surface
tension (reducing energy input from the initial rupture of
the surface) and, also, reduces the amplitude of thermal
fluctuations, thus suppressing
the formation of \org{multiple} thermal bridges \cite{perumanath2019droplet}.
Second, the presence of aggregates in the bulk hinders the 
flow of the water molecules and disperses the momentum 
transfer before it enters deeper into the droplets.

\section{Conclusions}
\label{conclusions}
In this study, the fundamental
processes involved in coalescence of droplets
containing surfactant have been described --- including
the initial rupture and bridge growth,
which occur on time and length scales 
inaccessible to experiment.
We have reported on
the main adsorption processes
(surfactant transport mechanism), 
characterised the bridge growth dynamics of
coalescence, and identified several important
differences to the case of pure water droplets
and those with surfactant.
Notably, Fig.~\ref{fig:5} suggests that if a slow-down of
coalescence processes is desired industrially, 
more surfactant should be added, which
confirms earlier 
suggestions\cite{jaensson2018tensiometry,leal2004flow}.
Moreover, we have identified early time collective
flow patterns that are present
in the case of aqueous droplets without surfactant, but 
are absent when appreciable surfactant is present. 
Surfactant also suppresses the multiple precursor
bridges that are important at early times for
pure water \cite{perumanath2019droplet}. 
The last appears to indicate that thermal
fluctuations will be less important for
topological changes of surfactant-laden 
droplets generally (splitting, merging, etc.).
We anticipate that our results open new exploration directions, 
which will be relevant for practical applications, 
and that they suggest the kind of effects 
that will be seen in other as yet 
unexplored processes such as 
droplet break-up and coalescence on substrates.
\blu{\org{An} 
aspect that requires further consideration
is the various effects that might be attributed to
a larger surface-area-to-volume ratio as the size
of the droplets decreases. For example, \org{we saw} 
that minor redistribution of surfactants from surface to bulk or vice
versa can cause large fluctuations in the bulk, while such
effects may be\org{come} negligible in macroscale systems.\cite{Politova2017} 
It would therefore be interesting to explore larger systems
in the future as more computational resources become available, 
as well as employ a range of different simulation models to explore
droplet coalescence in the presence of surfactant.}

\section{Supplementary Material}
\org{The} Supplementary Material provides the details of the 
probabilities for the mass transport mechanism of 
surfactant molecules between the different regions
in the droplets that reflect the arrows in
Fig.~\ref{fig:3}. It also contains data on
the bridge growth dynamics in the case of
C10E8 surfactant.

\begin{acknowledgments}
This research has been supported by the 
National Science Centre, Poland, under
grant No.\ 2019/34/E/ST3/00232. 
We gratefully acknowledge Polish high-performance 
computing infrastructure PLGrid (HPC Centers: ACK Cyfronet AGH) 
for providing computer facilities and support 
within computational grant no. PLG/2022/015261.
\end{acknowledgments}


\providecommand{\noopsort}[1]{}\providecommand{\singleletter}[1]{#1}%

\end{document}